\documentclass[a4paper,11pt]{article}
\pdfoutput=1
\usepackage{jheppub}
\usepackage[T1]{fontenc} 
\usepackage{amsmath,amssymb,graphicx,bbm,mathrsfs,nicefrac}
\usepackage{tikz}
\usepackage{amsmath}
\usepackage{amssymb}
\usepackage{graphicx,textcomp,float,gensymb,wrapfig, enumitem,comment,dsfont,framed,slashed,appendix,wrapfig,xcolor}
\usepackage{ bbold }
\usepackage{bm}
\usepackage{filecontents}
\usepackage{braket} 
\usepackage{hhline}
\usepackage{ mathrsfs }
\usepackage{rotating}
\usepackage[small]{caption}
\usepackage{subcaption}
\usepackage{etoolbox,hyperref,makecell}
\usepackage{feynmp}
\usepackage{accents}
\usetikzlibrary{arrows.meta}
\usepackage{empheq}
\usepackage{filecontents}

\newcommand{\stp}{X_{\tilde{t}}^0}
\newcommand{\mst}{m_{\tilde{t}}}

\usepackage{multirow}

\usepackage{braket} 

\patchcmd{\abstract}{\null\vfil}{}{}{}
\usepackage{pifont}

\newcommand{\beq}{\begin{equation}} \newcommand{\eeq}{\end{equation}}
\newcommand{\bea}{\begin{eqnarray}} \newcommand{\eea}{\end{eqnarray}}

\newcommand{\fref}[1]{Figure~\ref{#1}}

\newcommand{\be}{\begin{eqnarray}} \newcommand{\ee}{\end{eqnarray}}

\usepackage{hhline}
\usepackage{float}
\usepackage{caption}

\usepackage[utf8]{inputenc}

\preprint{FERMILAB-PUB-23-472}

\title{Di-Higgs Signatures in Neutral Naturalness}

\author[a]{Mario W. Barela,}
\author[b]{Rodolfo Capdevilla}

\affiliation[a]{Instituto de F\'\i sica Te\'orica, Universidade Estadual Paulista, R. Dr. Bento Teobaldo Ferraz 271 Barra Funda, S\~ao Paulo, SP 01140-070, Brazil}
\affiliation[b]{Theory Division, Fermi National Accelerator Laboratory, Batavia, IL 60510, USA} 

\emailAdd{mario.barela@unesp.br}
\emailAdd{rcapdevi@fnal.gov}

\date{\today}

\abstract{
The Higgs boson was the last fundamental piece of the Standard Model to be experimentally confirmed. LHC is embarked in a quest to probe the possibility that this particle provides a portal to new physics. One front of this quest consists in measuring the interactions of the Higgs with itself and with other SM particles to a high precision. In a more exotic front, the LHC is searching for the possibility that a pair of Higgses (HH) is the evidence of a new resonance. Such resonances are predicted in models with extended Higgs sectors, extra dimensions, and in models with exotic bound states. In this paper we show how scalar quirks in Folded Supersymmetry can give rise to HH resonances. We point out a viable sector of the parameter space in which HH is the dominant decay channel for these {\it squirkonium} bound states. We found that future runs of the LHC could discover HH resonances in the range of 0.5 - 1.6 TeV under reasonable assumptions. Furthermore, for a given mass and width of the HH signal, the model predicts the branching ratio of the subsequent decay modes of the heavy resonance. Finding the extra decay modes in the predicted pattern can serve as a smoking gun to confirm the model.
}

\begin{document}

\maketitle

\newpage
\section{Introduction}
\label{s.intro}

The current particle physics paradigm is that the Standard Model (SM) is a remarkable and, perhaps, the most successful existing physical theory. However, it is also known to be a low energy description of a much larger construction. This is because of the variety of phenomenological problems that the SM cannot address such as the Baryon asymmetry of the Universe, the mechanism for neutrino mass, flavor, and dark matter, to cite a few. One of the guiding principles in the search for physics beyond the SM has been Naturalness and the Hierarchy Problem (HP). This problem arises because the Higgs mass is quadratically sensitive to new physics scales, and becomes even more intriguing by the lack of evidence of new physics in ever increasing experimental energies. The SM is said unnatural for it does not contain a mechanism to stabilize the Higgs mass.

Solutions to the HP typically feature top partners responsible for cancelling the quadratic contribution to the Higgs mass from top quark loops. This is the case in the Minimal Supersymmetric version of the SM (MSSM). Unfortunately, the fact that the mass of the top partners has been pushed to an uncomfortably high regime by current data gives rise to a smaller leftover tuning referred to as {\it Little Hierarchy Problem}. It is the strong interacting quality of the top partners that results in the powerful constraints on their masses. This observation triggered the proposition of {\it Neutral Naturalness} \cite{Chacko:2005pe,Burdman:2006tz,Xu:2018ofw,Batell:2022pzc} models in which the top partners are neutral with respect to one or various of the subgroups of the SM group. Folded Supersymmetry (F-SUSY) is an example of this type of construction in which top partners are not charged under the SM QCD, but under a {\it dark} version of it. In this theory the Higgs mass is protected at the one loop level up to characteristic energies of tens of TeV. At this scale and above, it is possible to define an ultraviolet completion of F-SUSY with a fifth dimension compactified over an orbifold \cite{Burdman:2006tz}.

In F-SUSY the dark sector squarks are all heavier than the dark QCD hadronization scale. This causes them to behave as quirks (or squirks for its scalar nature). Pair production of these states results in excited {\it squirkonium} bound states that relax down to the ground state and decay promptly at collider time scales \cite{Burdman:2008ek}. Neutral squirkonium, here denoted as $X_{\tilde{q}}^0$, can be produced via $pp\to\gamma/Z\to\tilde{q}\tilde{q}^*$. Typically, these states preferentially decay into dark glueballs independently on the generation of the constituent squarks \cite{Curtin:2022tou,Albouy:2022cin,Curtin:2022oec,Batz:2023zef}. Charged squirkonium $X^+_{\tilde{q}}$, produced through $pp\to W\to \tilde{q}'\tilde{q}^*$, of the first and second generation will have a dominant branching ratio (BR) to $W+\gamma$ \cite{Harnik:2011mv,Burdman:2014zta,Capdevilla:2019zbx}. Now, third-generation charged squirkonium will undergo beta decay in a time scale much faster than relaxation \cite{Burdman:2008ek}, causing the system to decay to $W+X_{\tilde{q}}^0$, where $q$ represents the lighter between stop and sbottom. This final state shows promising results in a variation of the model where $X_{\tilde{q}}^0$ is longed-lived \cite{Li:2017xyf}.

F-SUSY production of third generation squirks always derives in neutral squirkonium, either by direct production or via beta decay of charged ones. This neutral state then preferentially decays to dark glueballs. One feature of the model is that the $0^{++}$ dark glueball state can mix with the Higgs boson through loops \cite{Juknevich:2009ji,Juknevich:2009gg}. This mixing causes the dark glueballs to have a naturally small coupling to SM particles, making them long-lived and a great signal for neutral naturalness models \cite{Curtin:2015fna,Chacko:2015fbc,Burdman:2018ehe}. However, glueball production is known to decrease as the mass splitting between the two stop eigenstates increases \cite{Chacko:2015fbc}. This is the regime that we will explore in this paper. We will see how increasing the soft trilinear term $A_t \tilde{t}_L \tilde{t}_R H$ that controls the mixing of the two eigenstops, causes the neutral {\it stoponium} state $X_{\tilde{t}}^0$ to predominantly decay to a pair of Higgs bosons. 

A similar observation was made long ago in the context of the MSSM, where studies of stoponium bound states \cite{Drees:1993yr,Drees:1993uw,Martin:2009dj,Kim:2014yaa,Kumar:2014bca,Batell:2015zla,Ito:2016qsm,Kang:2016wqi,Bodwin:2016whr,Duan:2017zar} have shown that Higgs decay modes dominate for large stop mixing angles. However, stoponium bound states can only be realized in the MSSM for low stop masses, in a regime excluded by the LHC. Our study brings back the possibility that HH resonances have a connection with the third generation of (s)quarks and Naturalness. Furthermore, we will see how the prediction of the model lies in a range of masses that will be soon explored by the LHC.

This paper is organized as follows: Sec. II gives a brief summary of the model and its unique phenomenological features. Sec. III presents our parametric setting where we define the benchmarks that we will analyze. We also show the theoretical bounds on the parameter space of interest from perturbative unitarity. Sec. IV shows squirkonium production cross section and decay modes. In Sec. V one can find our results for observability of HH resonances at the LHC. Finally, Sec. VI shows our conclusions and discussion.

\section{Scalar Quirks in Folded SUSY}
\label{s.modelindependent}

In this section we provide a synthesis of F-SUSY concepts that are important for our our analysis. For a complete treatment of the model, including a description of the full supersymmetric ultraviolet completion, we refer the reader to \cite{Burdman:2006tz}. In F-SUSY, the low energy theory is symmetric under the group $SU(3)_c \times SU(3)_{c^\prime} \times SU(2)_L \times U(1)_Y$. The representation content is that of the MSSM, but with squarks charged not under $SU(3)_c$, but under the \textit{dark} color $SU(3)_{c^\prime}$. The model comprises an additional octet of gluons corresponding to the new color sector.

In order to understand the origin of the strange dynamics this results in, it must be known that the two strong force groups are related to each other in the ultraviolet completion of the theory by a $Z_2$ symmetry. This ensures that the theory is fully Supersymmetric in the UV. As a consequence, the characteristic scales where confinement dynamics kicks in are close to each other $\Lambda_{c^\prime}  \sim \Lambda_c$.
In general, a pair-produced particle-antiparticle system will hadronize when the energy density of the flux tube (or string) approaches or exceeds $2 m_1$, where $m_1$ is the lightest quark-like particle in the theory.
Differently from QCD, the QCD$'$ particle content does not comprise any species with a mass $m$ smaller than the typical string tension $\Lambda_{c^\prime} $. Because of this, pair creation from the vacuum is suppressed as $\exp(- m_1^2/\Lambda^{\prime 2})$ and a produced pair of QCD$'$ particles will form a bound state instead of hadronizing. For this odd behavior, particles with charges of a strong group whose confining scale is much smaller than the lightest charged species mass are called \textit{quirks} \cite{Kang:2008ea,Cai:2008au,Kribs:2009fy,Craig:2015pha,Craig:2016kue,Knapen:2017kly,Li:2021tsy} -- and, in F-SUSY, since they are supersymmetric partners, \textit{squirks}.

At LHC energies and for lightest quirk masses of up to $\sim$ 1 TeV, the squirkonium will typically be produced at a highly excited state. A semiclassical analysis \cite{Burdman:2008ek} of the strong force bound state shows that the probability of decay only become appreciable after relaxation, \textit{i.e.}, after the excess energy is radiated away through emission of photons or glueballs, and the 2-particle system is left at the lowest lying angular momentum state. The decay of the squirkonium to lightest states will, then, most likely have an $s$ wave contribution. The possibility of detecting the soft signals of the relaxation period have been discussed in \cite{Harnik:2008ax} where the {\it antenna} pattern is the smoking gun signature.

Soon after the proposal of F-SUSY, the same authors showed that the $W+\gamma$ final state is the dominant decay mode for first and second generation of squirks. They also show that it is not possible to have a charged squirkonium bound state of the third generation because the heavier constituent will beta-decay in a timescale faster than relaxation \cite{Burdman:2008ek}. This indicates that only neutral squirkonium of the third generation is possible, a state which preferentially decays to dark glueballs. Now, the third generation is of great important for it is the one intrinsically tied to Naturalness and the hierarchy problem. Our work is motivated by this connection, and we would like to study decay channels of the neutral third-generation squirkonium in F-SUSY beyond those explored in the literature where long-lived glueballs seems to be one of the most interesting signals \cite{Curtin:2015fna}.

We will study the large soft trilinear coupling limit for stoponium, where the decay mode to HH can dominate over glueball formation. Our study only involves interactions of the third generation quarks, squarks and of the Higgs and gauge bosons. We will not make any attempt to fix classical problems of the MSSM like the $\mu$ problem or the Higgs mass \cite{Batra:2003nj,Draper:2013oza,Capdevilla:2015qwa,R:2021bml}. Our simplified analysis assumes: 1) The lightest stop is the lightest third generation squirk; 2) A neutral stoponium is produced from proton-proton collision at the LHC; 3) This state, initially highly excited, will promptly radiate away energy and angular momentum relaxing down to its ground state; 4) Finally, this ground state squirkonium will decay to a variety of channels with a narrow total width (below $5\%$). In order to determine if one of these channels can overcome glueball formation, we calculate the complete set of branching ratios and analyze their variation over an interesting sector of parameter space. We now discuss the parameter space of interest in the next section.

Before delving into the parameter space of interest in the next section, it is essential to underscore the distinctions between our simplified model and the one presented in the original F-SUSY paper and some of its pioneer phenomenological studies. While the original papers does not specifically emphasize the $A$-terms, we contend that, akin to any other soft term, these terms are generally viable in low-energy effective SUSY theories. We maintain an agnostic stance regarding the mechanism through which supersymmetry is broken, resulting in no specific prediction or ``natural'' choice for these parameters, apart from the intuitive $A_t \sim \Lambda_{\rm SUSY}$ (the scale of SUSY breaking). Furthermore, as of our current understanding, the constraints on large $A$ parameters stem mainly from phenomenological considerations and perturbativity, factors that we will duly incorporate into our analysis.
Concerning the low-energy spectrum, our theory deviates from the original papers solely in the hierarchy of squirk masses: ours is inverted, with the third generation being the lightest. This inversion is permitted by the parametric structure of F-SUSY.

\section{Parameter space and Constraints}
\label{s.singletmodels}

The interactions relevant to our study involve third generation squarks, gauge bosons, and the Higgs. These comprise, in principle, the following free parameters $$\{ \tan \beta, \: \mu, \: A_t, \: A_b, \: m_{\widetilde{Q}_L}, \: m_{\tilde{t}_R}, \: m_{\tilde{b}_R}\},$$ where $\tan \beta$ (or simply $t_\beta$) is the ratio $v_u/v_d$ of the vacuum expectation values (vev) of the two Higgses in the model, $\mu$ is the parameter of the supersymmetric quadratic scalar term, $A_q$ are the soft trilinear terms of the form $A_q H\widetilde{Q}_L \widetilde{q}_R$, and $m_{\widetilde{Q}_L}, m_{\tilde{t}_R}, m_{\tilde{b}_R}$ are the squark soft masses.

In order to define practical benchmarks, we choose a scenario in which all soft masses are equal \textit{i.e.}, $m_{\widetilde{Q}_L} = m_{\tilde{t}_R} = m_{\tilde{b}_R} \equiv \widetilde{m}_\text{soft}$ and there is no mixing in the sbottom sector, meaning $m_{\tilde{b}_{1}} =m_{\tilde{b}_{2}} = \widetilde{m}_\text{soft}$. These choices leave us with the following set of free parameters
\begin{equation}
\{ t_\beta, \: \mu, \: A_t, \: m_{\tilde{t}_1} \},
\end{equation}
where $m_{\tilde{t}_1}$ (or simply $m_{\tilde{t}}$) is the mass of the lightest eigenstop. A given choice of these parameters will determine the mass of the heaviest stop, the soft (and sbottom) mass, and mixing angles.

In our analysis, we will vary the mass of the lightest stop between 250 GeV up to 1 TeV and the soft trilinear parameter from 1 up to a few TeV. It could be argued that a {\it natural} choice for the other parameters is $(t_\beta, \mu)\sim(1, m_h)$, where $m_h$ is the mass of the SM-like Higgs particle. A {\it tuned} choice of $(t_\beta, \mu)$ could be defined as one that reflects a hierarchy between the two vev of the model and between $\mu$ and the EW scale. Without a rigorous definition of tuning, here we define a set of benchmarks (B1, B2, B3, B4) that go from very small to some degree of tuning:
\begin{equation}
\begin{split}
{\rm B1}\hspace{-0.1cm}:& \quad \mu = 200\,{\rm GeV}, \, t_\beta = 1\\
{\rm B2}\hspace{-0.1cm}:& \quad \mu = 200\,{\rm GeV}, \, t_\beta = 10\\
{\rm B3}\hspace{-0.1cm}:& \quad \mu = 1\,{\rm TeV}, \, t_\beta = 1\\
{\rm B4}\hspace{-0.1cm}:& \quad \mu = 1\,{\rm TeV}, \, t_\beta = 10.
\end{split}
\label{eq:bench}
\end{equation}

As mentioned above, $A_t$ is the scalar trilinear coupling that controls the $H\tilde{t}_{1}\tilde{t}_{1}^{*}$ vertex strength. Increasing this parameter increases the splitting between the two eigenstops $\tilde{t}_1, \tilde{t}_2$ which, as we will see below, in turn increases the production and HH decay rates of the squirkonium states of interest. However, a trilinear term like $A_t$ cannot be set to arbitrarily large values because this parameters can create problems like violation of perturbative unitarity \cite{Schuessler:2007av,Goodsell:2018tti}, or it can induce large trilinear Higgs self-couplings that in turn can violate constraints from EW (S/T parameters, weak angle, etc.) and Higgs physics (Higgs branching ratios, double Higgs production, etc.) \cite{Peskin:1991sw,Kribs:2017znd,Degrassi:2021uik,Lu:2022bgw}. We now discuss the relevant constraints.

\subsection{Perturbative Unitarity}

To verify the validity of the theory we now study partial wave unitarity in the parameter space of interest. We begin from the partial wave expansion of the (azimuthally symmetric) scattering amplitude for the scalar $2\to2$ process ${i \to f \equiv} \{a,b\}\to\{c,d\}$, here denoted by $\mathcal{M}_{if}(\theta)$. The $j$-th coefficient of the expansion is
\begin{equation}
\label{e.pwd}
a_{if}^{j}=\frac{1}{32\pi}\sqrt{\frac{4|\textbf{p}^i||\textbf{p}^f|}{2^{\delta_{ab}}2^{\delta_{cd}}}}\int d\theta \mathcal{M}_{if}(\theta) P_j(\theta),
\end{equation}
where $P_j(\theta)$ are the Legendre polynomials and $\textbf{p}^i,\textbf{p}^f$ are the centre of mass three-momentum for the initial and final states respectively. In a multi-process analysis one can construct the matrix $(a^{j=0})_{if}$ taking into account all the initial and final states. To satisfy the unitarity condition, the $k$-th eigenvalue of this matrix must obey
\begin{equation}
\left|\text{Re}\left(a^k_0\right)\right| \leq \frac{1}{2}, \;\; \forall \, k.
\end{equation}
Note that the constraint above must hold in the entire phase space. To obtain an estimate of the unitarity bounds, we consider the amplitude for the process $\tilde{t}_1 \tilde{t}_1^* \to \tilde{t}_1 \tilde{t}_1^*$, which include the $4$-scalar vertex as well as $s$- and $t$-channel exchange of Higgs and dark gluons. The $0$-th coefficient is given by\footnote{These approximate formulae ignore terms proportional to EW parameters suppressed by factors of $m_Z^2/m_t^2$ and $m_Z^2/m_{\widetilde{t}_1}^2$. In our analysis and figures no approximations have been considered.}
\begin{equation}
a_0\sim-\frac{1}{24  \pi s_\beta^2 v_h^2} \sqrt{1 - \frac{4 m_{\tilde{t}}^2}{s}} (F_0 + F1 + F_2 + F_3),
\label{eq:a0}
\end{equation}
where
\begin{equation}
\begin{split}
F_0=& (3 m_t^2 s_{2\theta}^2 + g_s^2 s_\beta^2 c_{2\theta}^2 v_h^2)\\
F_1=& e^2s_\beta^2(9 c_\theta^4 + 8s_W^2 (2s_\theta^4-c_\theta^2))/(12c_W^2s_W^2)\\
F_2=& \frac{6 m_t^2 (c_\alpha m_t + s_\theta c_\theta (A_t c_\alpha - s_\alpha \mu))^2}{
 s - m_h^2}\\
F_3=& -\frac{s - m_h^2}{s - 4 m_{\tilde{t}}^2} F_2 \log\left[1 + \frac{s - 4 m_{\tilde{t}}^2}{m_h^2}\right].
\end{split}
\end{equation}
Here, $m_t$ is the mass of the top quark, $v_h$ is the SM-like Higgs vev, $\theta$ is the stop mixing angle, and $\alpha/\beta$ are the mixing angles of the neutral CP-even/odd components of the two Higgs multiplets in the MSSM \cite{Martin:1997ns}.

\begin{figure}[t!]
\centering
\includegraphics[width=0.5\linewidth]{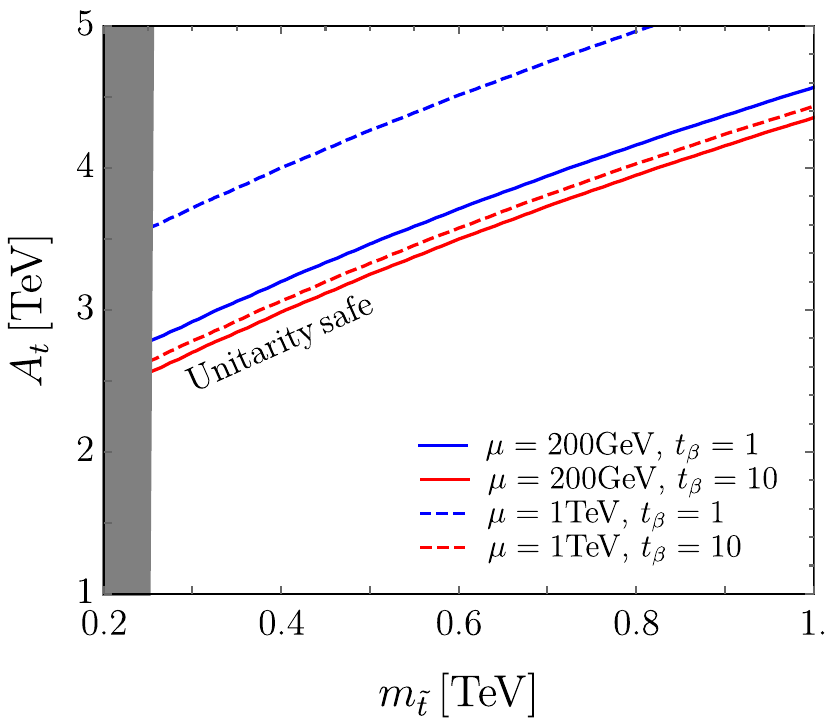}
\caption{Maximum $A_t$ allowed by perturbative unitarity as a function of the lightest stop mass.}
\label{fig:unit}
\end{figure}


Fig.~\ref{fig:unit} shows the unitarity bounds corresponding to our four benchmarks defined in Eqs. \ref{eq:bench}. Below each line the model is unitary safe. We found that for a stop mass of 250~GeV the bound on $A_t$ varies between 2.5 and 3.5~TeV, depending on the benchmark. Note how reducing $\mu$ and increasing $\tan\beta$ one may extend the allowed region of parameter space. For a more refined calculation, one can construct a $5\times5$ scattering matrix including $hh,\, \tilde{t}_1 \tilde{t}_1^*,\, \tilde{t}_2 \tilde{t}_2^*,\, \tilde{b}_1 \tilde{b}_1^*,\, \tilde{b}_2 \tilde{b}_2^*$ initial and final states. In \cite{Schuessler:2007av} the authors show how including some of these processes one can extend the unitary bound on $A_t$ up to 4.4 - 5~TeV for stop masses of 100~GeV. We will keep our calculation as a conservative constraint keeping in mind that the full calculation could in principle open a larger region of parameter space.

\subsection{Higgs Physics Constraints}

The interaction $A_t H\tilde{t}_i\tilde{t}_i^{*}$ can generate a loop level correction to the Higgs self-coupling. We calculated this correction in the broken phase where loops involving the eigenstops $\tilde{t}_{1,2}$ generate three-Higgs interactions. The full formula is a complicated expression of the parameters of the model, but for illustration we provide here the dominant term in the limit where both the stop masses and the trilinear term $A_t$ are much larger than the EW scale
\begin{equation}
\delta\lambda_{3h} \sim \frac{c_\alpha^3s_{2\theta}^3y_t^3}{\sqrt{2}\,64\pi^2}\frac{A_t^3}{m_{\tilde{t}}^2}.
\end{equation}
This correction can be dramatically large as the parameter $A_t$ grows for a given spectrum of Folded stops $\tilde{t}_i$. As we will see later, for a given $t_\beta$, increasing $\mu$ is a way to reduce the size of the correction (keeping in mind that the full expression is more complicated polynomial of the quantity $A_t$ and not just it to the third power).

To set constraints, we follow the results from \cite{Degrassi:2021uik} where the authors find constraints on new physics corrections to the Higgs self-coupling from a variety of measurements including the $W$ mass, the weak angle, EW precision observables, Higgs boson analysis in the $\gamma\gamma, ZZ^*, WW^*, \tau\tau$ and $b\bar{b}$ channels, as well as double Higgs productions analysis. As we will see later, these constraints have a dramatic impact on the {\it natural} benchmarks defined above, and they can only be ameliorated as one moves in the parameter space towards the {\it tuned} benchmark points.

While constraints from large contributions to $\delta\lambda_{3h}$ have the more severe impact on our parameter space, for completeness, we also looked into the constraints from $h\to\gamma\gamma$. Large contributions to this decay width $\delta\Gamma_{h\gamma\gamma}$ can be induced by loops of stops, which come proportional to two powers of $A_t$. We found that this process can constrain a small portion of the parameter space in the low mass regime not covered by constraints from $\delta\lambda_{3h}$ as we will see later in Sec.~\ref{s.implications}.

\subsection{Electroweak Oblique Parameters}

Our parameter space of interest includes relatively light EW states. These are known to add large contributions to the EW oblique parameters $S, T,$ and $U$~\cite{Peskin:1991sw}. We calculated the contributions to these parameters in our simplified model finding
\begin{equation}
S=\frac{1}{12\pi\Delta_{21}^{3}}\left[S_{0}+S_{1}\log\left(\frac{m_{\tilde{t}_{2}}^{2}}{m_{\tilde{t}}^{2}}\right)+S_{2}\log\left(\frac{m_{\tilde{b}}^{2}}{m_{\tilde{t}}^{2}}\right)\right],
\end{equation}
\begin{equation}
T=\frac{1}{16\pi s_{W}^{2}c_{W}^{2}m_{Z}^{2}}\left[T_{0}+T_{1}\log\left(\frac{m_{\tilde{t}_{2}}^{2}}{m_{\tilde{t}}^{2}}\right)+T_{2}\log\left(\frac{m_{\tilde{b}}^{2}}{m_{\tilde{t}}^{2}}\right)+T_{3}\log\left(\frac{m_{\tilde{t}_{2}}^{2}}{m_{\tilde{b}}^{2}}\right)\right],
\end{equation}
where\footnote{The expression for $U$ is quite lengthy, very little illuminating, and we verified that $U\ll S,T$ in our parameter space of interest.}
\begin{equation}
\begin{split}
\Delta_{21} & =  m_{\tilde{t}_{2}}^{2}-m_{\tilde{t}}^{2}, \\
S_{0} & = -s_{\theta}^{2}c_{\theta}^{2}\Delta_{21}(5m_{\tilde{t}}^{4}-22m_{\tilde{t}_{2}}^{2}m_{\tilde{t}}^{2}+5m_{\tilde{t}_{2}}^{2}), \\
S_{1} & = s_{\theta}^{2}(2c_{\theta}^{2}-s_{\theta}^{2})\Delta_{21}^{3}-6s_{\theta}^{2}c_{\theta}^{2}m_{\tilde{t}}^{4}(3m_{\tilde{t}_{2}}^{2}-m_{\tilde{t}}^{2}), \\
S_{2} & = \Delta_{21}^{3}, \\
T_{0} & = 3(s_{\theta}^{4}m_{\tilde{t}_{2}}^{2}+m_{\tilde{b}}^{2}+c_{\theta}^{4}m_{\tilde{t}}^{2}),\\
T_{1} & = \frac{6s_{\theta}^{2}c_{\theta}^{2}m_{\tilde{t}}^{2}(m_{\tilde{t}}^{2}+m_{\tilde{t}_{2}}^{2})}{\Delta_{21}},\\
T_{2} & = -\frac{6c_{\theta}^{2}m_{\tilde{b}}^{2}m_{\tilde{t}}^{2}}{m_{\tilde{b}}^{2}-m_{\tilde{t}}^{2}},\\
T_{3} & = -\frac{6s_{\theta}^{2}m_{\tilde{b}}^{2}m_{\tilde{t}_{2}}^{2}}{m_{\tilde{t}_{2}}^{2}-m_{\tilde{b}}^{2}}.
\end{split}
\end{equation}
To set constraints, we will follow~\cite{Lu:2022bgw} and draw the $2\sigma$ contours corresponding to the PDG2021 data fit. The main constrain comes from the $T$ parameter and we will show in the results in Sec.~\ref{s.implications} how these constraints also cover a small portion of the parameter space in the low mass regime not covered by constraints from $\delta\lambda_{3h}$. We note how for a given $\mst$, as one increases $A_t$ it increases the splitting between $\tilde{t}$ and the heavier states, increasing the impact on the EW oblique parameters, giving stronger constraints.


Before proceeding with the phenomenology of the model, we argue that our choice $\mst>$250~GeV serves as a conservative lower bound on the mass of the lightest stop. We point out how recent works consider a similar lower bound on 3rd generation folded stop masses~\cite{Batz:2023zef} which comes from the possibility that the Higgs decays to a pair of glueballs adding an extra contribution to the invisible branching ratio of the Higgs~\cite{Chacko:2015fbc}. Stronger lower bounds on folded squirk masses can be found for the 1st and 2nd generation squirks $m_{\tilde q}>800$~GeV~\cite{Capdevilla:2019zbx} whose main decay channel to $W+\gamma$ makes it a state easy to constrain. For 3rd generation squirks, in the limit of heavy glueball masses ($\sim40$~GeV) and small mixing in the stop sector, the lower bound on $\mst$ can be as strong as 1 TeV, but in the limit of small mixing and light glueballs the bounds are quite weak~\cite{Chacko:2015fbc}.

\section{Stoponium Production and Decay}
\label{s.ewmodels}

\subsection{Production}

We now discuss the production mechanisms for our squirkonium state of interest at the LHC. In the parameter space that we focus i.e. where the trilinear term $A_t$ is large, the dominant production channels of stoponium $\stp$ are
\begin{equation*}
\begin{split}
q\bar{q}\, {\rm fusion}: & \quad p(q)p(\bar{q})\to\gamma/Z\to\tilde{t}\tilde{t}^*\\
gg\, {\rm fusion}: & \quad p(g)p(g)\to h\to\tilde{t}\tilde{t}^*.
\end{split}
\end{equation*}
The first process is the usual Drell-Yan, neutral gauge boson mediated, $q\bar{q}$-fusion. The second process is the $gg$-fusion that involves a triangle top-quark loop and a Higgs in the $s$-channel. In the limit of large $A_t$ and high center of mass energy, the partonic cross section of the $q\bar{q}$-fusion is given by
\begin{equation}
\hat{  \sigma}(q\bar{q}\to \tilde{t}\tilde{t}^*)\approx \frac{\pi \alpha^2}{3\hat{s}}\left(1-\frac{4m_{\tilde{t}_1}^2}{\hat{s}}\right)^{3/2}f_q(\theta),
\end{equation}
where $f_q(\theta)=\alpha^q_0+\alpha^q_2s_\theta^s+\alpha^q_4s_\theta^4$. The dimensionless coefficients $\alpha^q_i$ are given in terms of SM constants and are numerically equal to $\alpha_0^u=20.3, \alpha_2^u=-32.8, \alpha_4^u=18.2$, and $\alpha_0^d=17.6, \alpha_2^d=-39.3, \alpha_4^d=23.4$. In the same limit of large $A_t$ and $\hat{s}$, the partonic cross section of the $gg$-fusion process is given by
\begin{equation}
\hat{  \sigma}(gg\to \tilde{t}\tilde{t}^*)\approx\frac{6 \alpha_s^2y_t^2m_t^4}{64^2\pi^3\hat{s}^2v_h^2}\left(1-\frac{4m_{\tilde{t}_1}^2}{\hat{s}}\right)^{1/2}g_t(\hat{s}),
\end{equation}
\vspace{-0.3cm}
\begin{equation}
g_t=\frac{s_{2\theta}^2A_t^2}{4t_\alpha^2\hat{s}}\left[-4+\left(1-\frac{4m_t^2}{\hat{s}}\right)\log^2\left(-\frac{m_t^2}{\hat{s}}\right)\right]^{2}.
\end{equation}
In our calculation we included the effects of $u,d,s,c,g$ partons convoluting the cross section above with the corresponding PDFs for which we used the MSTW2008 set \cite{Martin:2009iq}.

\begin{figure*}
	\centering
\includegraphics[width=0.5\linewidth]{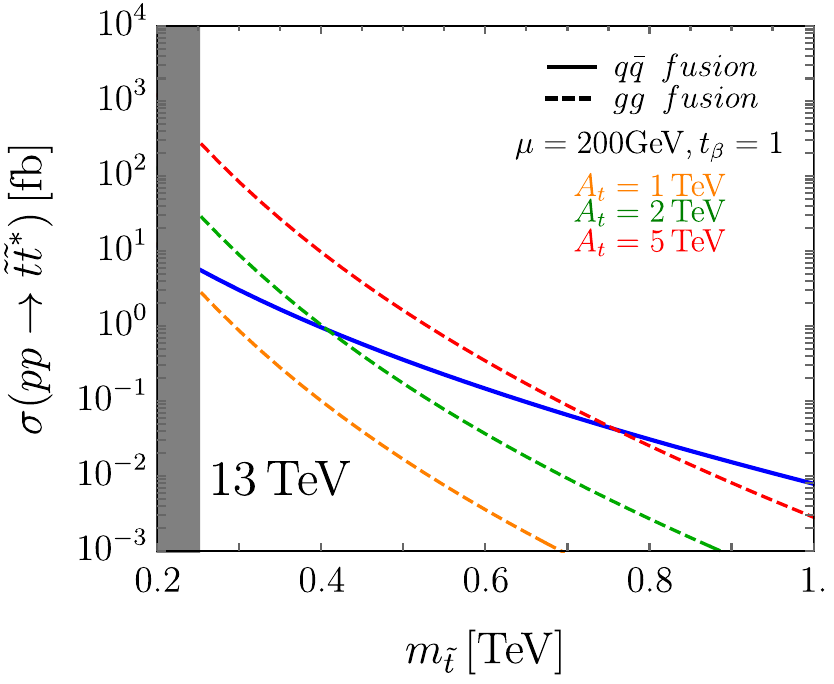}~~~\includegraphics[width=0.5\linewidth]{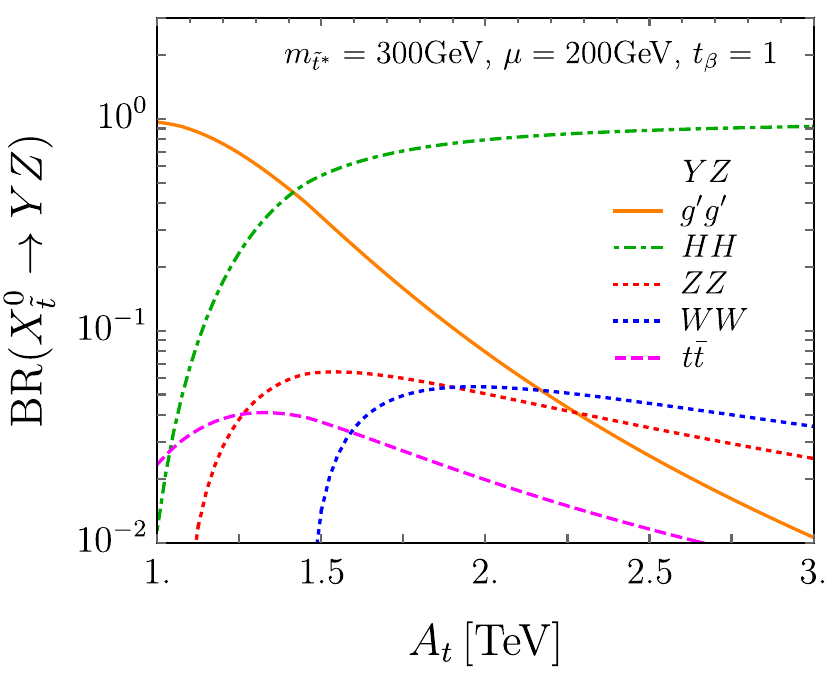}
		\caption{Left: Production cross section of stoponium at the LHC. For low $A_t$ values, the dominant process is $q\bar{q}$-fusion, whereas $gg$-fusion dominates for large $A_t$. Right: Branching Ratios of the lowest lying energy state of the lightest stoponium into the various decay modes as a function of $A_t$.}
	\label{fig:xsec}
\end{figure*}

The cross sections resulting from these channels may be observed in Fig.~\ref{fig:xsec} (left). The $q\bar{q}$-fusion process (solid blue) occurs through gauge interactions and it is independent of $A_t$. The $gg$-fusion channel (dashed lines) involves a $H\tilde{t}_1\tilde{t}^*_1$ vertex and it is enhanced with increasing $A_t$, reason why this channel dominates for an arbitrarily high value of this parameter. Note, for example, that for a mass of $m_{\tilde{t}_1} =  0.4$ TeV the $gg$-fusion process dominates for $A_t >  2$ TeV.

\subsection{Decay}

In order to calculate the BR of the different decay modes of $\stp$ we will follow the method in \cite{Burdman:2008ek}. We calculate the cross section $  \sigma(\tilde{t}\tilde{t}^*\to xy)$ for all possible combinations of $xy$ given the interactions of the $\stp$ state: $g'g', HH, H\gamma, HZ, \gamma\gamma, \gamma Z, ZZ, WW, t\bar{t}$. We then get the annihilation rate $\langle   \sigma v \rangle$ taking the limit where the relative velocity $v$ of the $\tilde{t}\tilde{t}^*$ system goes to zero. Finally, the BR for the $i$-th decay mode is simply ${\rm BR}_i = \langle   \sigma v \rangle_i/\sum_j \left\langle   \sigma v \right\rangle_j$.

A priori, one can guess that the dominant decay mode is $g'g'$ due to strong nature of the interaction. Our task is to look for a region of the parameter space where HH can dominate. In the limit $A_t\gg m_{\tilde{t}} \gg m_t, m_h$, the $g'g'$ and HH annihilation rates are equal to
\begin{equation}
\begin{split}
\langle   \sigma v \rangle_{g'g'} \approx & \frac{28\pi \alpha_s^2}{3m_{\tilde{t}}^2},\\
\langle   \sigma v \rangle_{HH} \approx & \frac{3y_t^4c_\alpha^4s_{2\theta}^4A_t^4}{128\pi m_{\tilde{t}}^6}.
\end{split}
\end{equation}
Here we can observe that for large enough values of $A_t$, the HH mode is expected to dominate\footnote{It is crucial to consider that, for sufficiently large glueball masses and depending on the matter content above the Electroweak (EW) scale, the dark coupling constant $\alpha_{s'}$ can undergo significant running, becoming much larger than the QCD coupling, as discussed in \cite{Curtin:2015fna}. In principle, this possibility could make it substantially more challenging for the HH mode to dominate the squirkonium branching ratio (BR) pattern. In our model, we assume that the glueball mass is small enough so that $\alpha_{s'}$ does not significantly deviate from $\alpha_{s}$.}. In agreement with this intuition we can see in Fig. \ref{fig:xsec} (right) how for large $A_t$, the $g'g'$ mode (solid orange) is highly suppressed whereas the HH mode (dot-dashed green) BR approaches one. The effect of increasing the stop mass $m_{\tilde t}$ (not shown in the figure) is that all curves in the figure move to the right, meaning that the HH mode starts dominating at higher values of $A_t$ than those shown in the figure. In the relevant parameter space, we found that the modes $H\gamma, HZ, \gamma\gamma, \gamma Z$ were highly suppressed compared to those shown in Fig. \ref{fig:xsec}.

\section{Di-Higgs Signals at the LHC}
\label{s.implications}

The LHC performs both resonant and non-resonant searches for a pair of Higgs bosons in a variety of final states \cite{CMS:2017hea,ATLAS:2018dpp,ATLAS:2018uni,CMS:2018ipl,ATLAS:2019qdc,DiMicco:2019ngk,CMS:2021roc,ATLAS:2021ifb,ATLAS:2022hwc,CMS:2022kdx,ATLAS:2022xzm,ATLAS:2022fpx}. One of the main motivations of HH searches is to accurately measure the self coupling of the Higgs. The SM has an unfortunate accidental cancellation between the two main diagrams that contribute to HH production, namely, the gluon fusion s-channel Higgs exchange that then splits into two Higgses via self coupling, and the gluon fusion to HH via a top quark box diagram. The total cross section for this process in the SM is about 33.47 fb \cite{Spira:2016zna,Borowka:2016ypz,Cao:2016zob,Jones:2017giv,Davies:2018ood,Jones:2018hbb,Grazzini:2018bsd,Bonciani:2018omm,DeFlorian:2018eng,Dawson:2018dcd,Banerjee:2018lfq,Xu:2018eos,Basler:2018dac,Davies:2019dfy,Davies:2019djw,Baglio:2020ini,Amacker:2020bmn,Baglio:2020wgt,Wang:2020nnr,deFlorian:2021azd,Huang:2022rne,Alioli:2022dkj,Davies:2023npk,Davies:2022ram,Chen:2019lzz,Chen:2019fhs,AH:2022elh}. The main effect of the self coupling is more significant at lower HH invariant masses. Current bounds from non-resonant HH searches at the LHC constrain the trilinear coupling to be within 40\% of the SM prediction \cite{Kim:2018uty,Goncalves:2018qas,Heinrich:2019bkc,Agrawal:2019bpm,Heinrich:2020ytv,ATLAS:2022jtk,Arco:2022lai,Alasfar:2023xpc}.

Now, the fact that HH has a small cross section in the SM opens an opportunity for new physics. In the large invariant mass regime one expects very little irreducible background events. Searches for HH resonances performed in the $bbbb$ final states place bounds \cite{ATLAS:2022hwc} on masses between 250 GeV and 5 TeV for spin 0 \cite{Lewis:2017dme} and spin 2 \cite{Carvalho:2014lsg} resonances. The bounds on the cross section times HH branching ratio range between a few pb for the lowest masses down to 1 fb for the heaviest mass. \footnote{These bounds imply different lower bounds in the HH resonance mass in the context of different models \cite{Li:2019tfd,Chen:2018wjl,Cheung:2020xij,Abouabid:2021yvw,Anisha:2022hgv,Chung:2022kjp,Iguro:2022fel,Baglio:2023euv,Englert:2019eyl,Alasfar:2019pmn,Moretti:2023dlx,Gabriel:2023dyx,Banerjee:2016nzb}.}

\begin{figure*}
\centering
\includegraphics[width=0.5\textwidth]{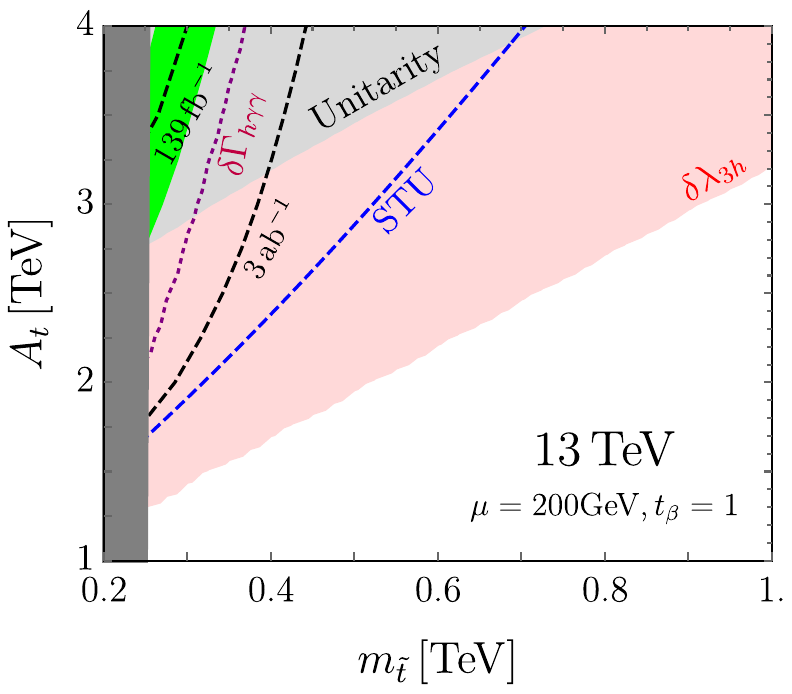}~\includegraphics[width=0.5\textwidth]{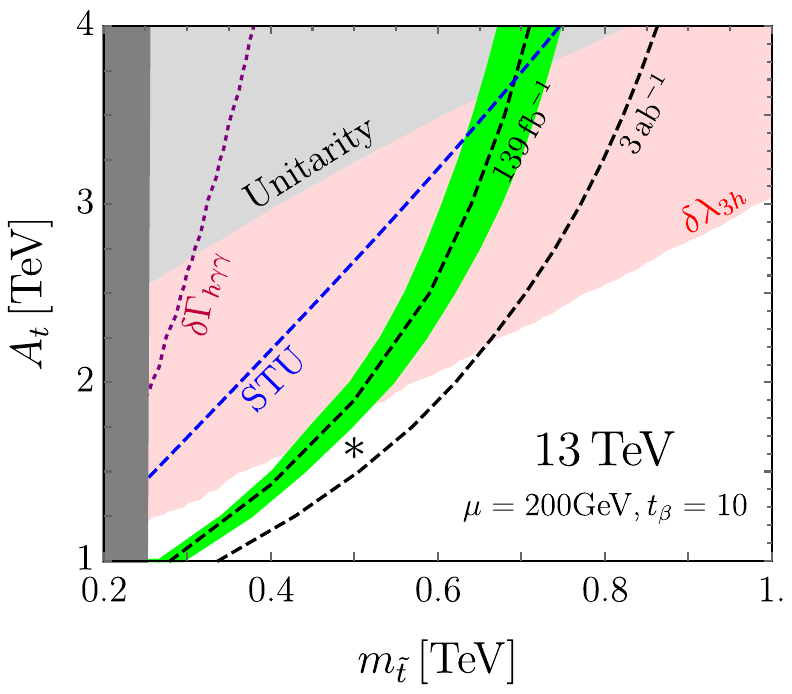}
\caption{Exclusion contours on the $(\mst, A_t)$ plane for the two {\it natural} benchmarks. For low $t_\beta$ the LHC is expected to find low mass resonances in the range of (500, 800) GeV corresponding to $\mst$ in the range (250, 400) GeV. This region of parameters is disfavored by the constraints we considered; unitarity (light gray), corrections to the trilinear Higgs coupling (light red), corrections to the Higgs width to a pair of photons (dotted purple), and EW oblique parameters (dashed blue). As $t_\beta$ increases, heavier resonances are expected so that for $t_\beta=10$ a 1.2 TeV resonance is possible, corresponding to $\mst=600$ GeV. There is a small window of masses, that evade constraints, where the HL-LHC could find HH resonances from this model as we can see at the bottom of the dashed lines. The black star represents the benchmark $\mst=500$ GeV and $A_t=1.63$ TeV.}
\label{fig:moneyplot1}
\end{figure*}

\begin{figure*}
\centering
\includegraphics[width=0.5\textwidth]{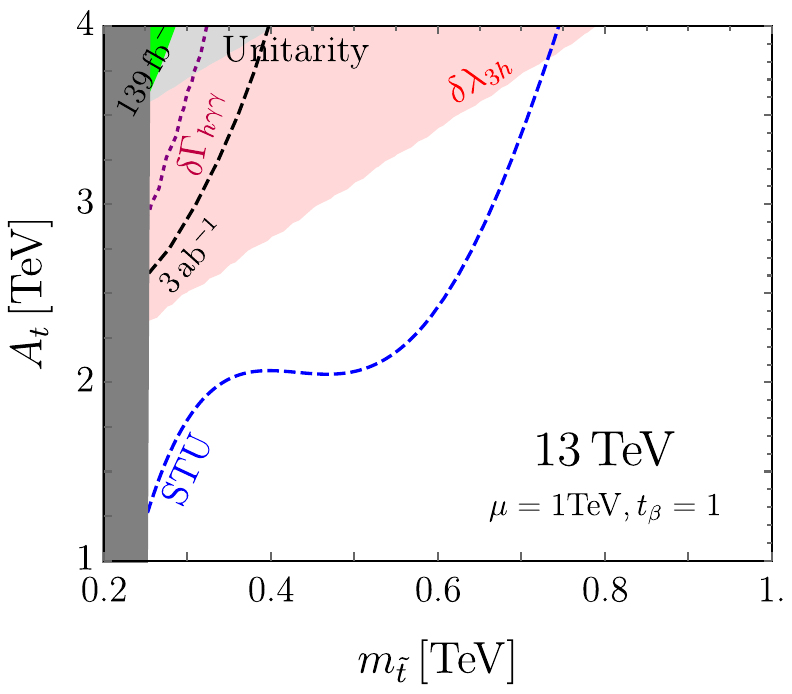}~\includegraphics[width=0.5\textwidth]{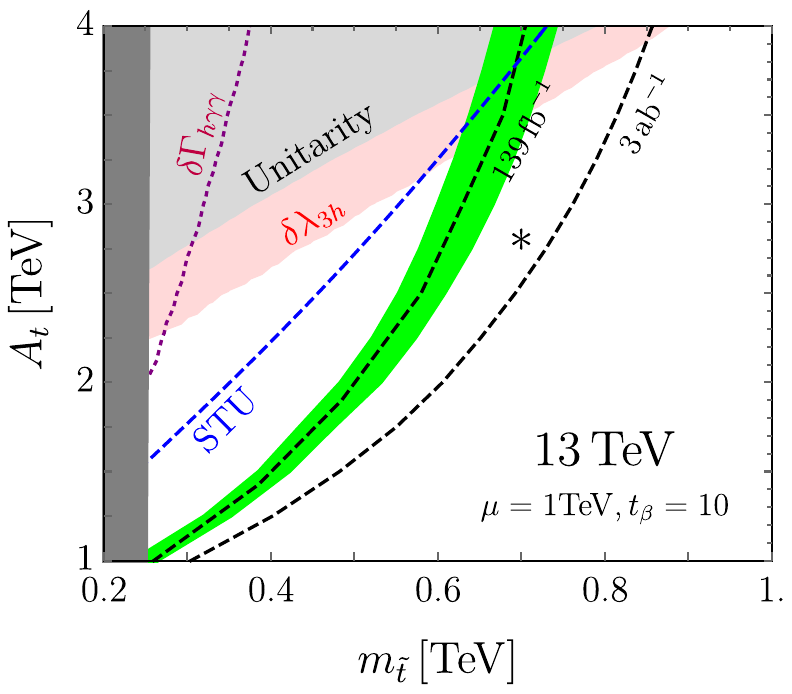}
\caption{Similar to \fref{fig:moneyplot1} but for the {\it tuned} benchmarks. The projected sensitivities are similar to those of the {\it natural} benchmarks because both production and decay of stoponium have a small dependence on $\mu$ in the parameter space of interest. However, the constraints ameliorate as $\mu$ increases, which implies a wider range of masses for which HL-LHC could find a HH resonance related to this model. The black star represents the benchmark $\mst=700$ GeV and $A_t=2.8$ TeV.}
\label{fig:moneyplot2}
\end{figure*}

In order to find the reach of the LHC on the parameter space of our model, we calculated the cross section for stoponium production and multiplied by the corresponding BR to HH in the plane $(\mst, A_t)$. Our results are presented in Fig. \ref{fig:moneyplot1} and \ref{fig:moneyplot2}, where we show the exclusion and projections for the different benchmarks defined in Eq. \ref{eq:bench}. We found that for the {\it natural} benchmark B1, where $\mu = 200$ and $t_\beta=1$ (Fig. \ref{fig:moneyplot1} - left), both current LHC data and HL-LHC projections only probe a region of the parameter space that is disfavored by the constraints; where Unitarity corresponds to the shaded light-grey region, constraints from corrections to the Higgs trilinear coupling ($\delta\lambda_{3h}$) correspond to the red shaded region, and constraints from modifications to the Higgs to photon decay width $(\delta\Gamma_{h\gamma\gamma})$ and from the EW oblique parameters ($STU$) are denoted by the purple-dotted and blue-dashed lines, respectively. For the second {\it natural} benchmark B2, where $\mu = 200$ and $t_\beta=10$ (Fig. \ref{fig:moneyplot1} - right), current data exclude resonances up to 900 GeV, corresponding to stop masses of 450 GeV in a small corner of the parameter space outside the constraints (bottom-left corner of the plot). According to this benchmark, HL-LHC will discover HH resonances in the range $\sim 0.6 - 1.3$ TeV corresponding to stop masses between $\sim300-650$ GeV.

The bottom line of what these results indicate is that the LHC could discover di-Higgs resonances in the range of 600 - 1300 GeV in subsequent runs, as indicated by the unconstrained region between the green shaded band and the HL-LHC projection black dashed line. This, in a reasonable {\it natural} region of the parameter space. Furthermore, if LHC finds a HH resonance in this range, according to our analysis, we will be able to infer the value of $t_\beta$ and $A_t$ within a small window. This in turn will allow us to infer the subsequent decay modes of the resonance according to the right panel of Fig. \ref{fig:xsec}. As we can see in said figure, our resonance will have a significant BR to massive gauge bosons, and if this resonance were to be related to naturalness and the stops, it will also have a significant BR to a pair of top quarks. Finding the same resonance in any of these channels would amount to strong evidence in favour of the model.

The situation for the more {\it tuned} benchmarks B2 and B3, for which $\mu=1$ TeV (Fig. \ref{fig:moneyplot2}), is quite similar to what happens for the {\it natural} benchmarks; future runs of the LHC could discover HH resonances in the range of 600 - 1600 GeV as a function $t_\beta$. To clarify this point, we highlight the fact that moving from the left to the right panel of the figure, we can see how increasing the value of $t_\beta$ in the range between 1-10, the LHC data can probe resonances with masses of 500 GeV (corresponding to stop masses of 250 GeV) up to heavier resonances at 1.6 TeV (corresponding to stop masses of 800 GeV) in a region outside the constraints.

As discussed in previous sections, our calculations assume stoponium production, fast relaxation, prompt decay, and a narrow width so that our signal efficiency is comparable to those of the LHC searches. Except for the last, all these assumptions were proved to be valid for stoponium in Folded SUSY \cite{Burdman:2008ek}. Here, we would like to discuss further on the narrow width assumption. In the non-relativistic limit, the partial width for $X_{\tilde t}^0\to A+B$ is given by~\cite{Kuhn:1979bb,Guberina:1980dc,Barger:1987xg}
\begin{equation}
\Gamma(X_{\tilde{t}}^{0}\to A+B)=\frac{3\beta}{32\pi^{2}}\frac{|R(0)|^{2}}{(2m_{\tilde{t}})^{2}}\frac{1}{1+\delta_{AB}}|\mathcal{M}(\tilde{t}\tilde{t}^{*}\to AB)|_{v=0}^{2},
\end{equation}
where $\delta_{AB}$ is a statistical factor, the last term is the square of the matrix element of the process ${\tilde t}{\tilde t}^*\to AB$ evaluated at zero relative velocity, $|R(0)|$ is the wave function of the squirkonium bound state when the system collapses into a zero relative distance of the constituents, and $\beta$ is as usual
\begin{equation}
\beta=\sqrt{\left(1-\frac{m_{A}^{2}+m_{B}^{2}}{(2m_{\tilde{t}})^{2}}\right)^{2}-\frac{4m_{A}^{2}m_{B}^{2}}{(2m_{\tilde{t}})^{4}}}.
\end{equation}
We now make a rough estimate of the width of our neutral stoponium corresponding to the two benchmark points highlighted in the right panels of Figs.~\ref{fig:moneyplot1}~and~\ref{fig:moneyplot2} that we dub as ``signal benchmark $S_1$'' and ``signal benchmark $S_2$'' for which
\begin{equation*}
\begin{split}
S_1:\quad(\mst, A_t,\mu,t_\beta) & = (500\,{\rm GeV}, 1.63\,{\rm TeV}, 200\,{\rm GeV}, 10), \,\, {\rm and} \\
S_2:\quad(\mst, A_t,\mu,t_\beta) & =(700\,{\rm GeV}, 2.8\,{\rm TeV}, 1\,{\rm TeV}, 10),
\end{split}
\end{equation*}
respectively. To estimate $|R(0)|$ we use classic results on heavy quarkonium bound states (see e.g.~\cite{Hagiwara:1990sq}), where we have $|R(0)|^2/(2\mst)^2\sim0.2$ GeV for a mass of $\mst=500$ GeV. We can see in Table~\ref{tab:BR} a comparison of the BR of stoponium to a pair of dark gluons and a pair of Higgses and the corresponding decay width. We see that in the first signal benchmark the $A_t$ term is not large enough to make the HH channel dominate over $g'g'$. In this case the partial width $\Gamma_{\rm HH}$ is about a factor of two smaller than $\Gamma_{g'g'}$. For the second signal benchmark, the large $A_t$ term causes the partial width $\Gamma_{\rm HH}$ to be almost a factor of two larger than $\Gamma_{g'g'}$, which remains constant as it does not depend on $A_t$. These sub GeV partial widths are in the ballpark of those reported in classic stoponium literature e.g.~\cite{Hagiwara:1990sq} where we can find for example that the $0^{-+}$ state, has a total decay width of about 1 GeV for $\mst=500$ GeV, including the channels $gg,\gamma\gamma,f{\bar f},Z\gamma,ZZ,WW,$ and $ZH$. Although our estimates in Table~\ref{tab:BR} justify our ``narrow width'' assumption, we believe that a more thorough investigation of the width in the large $A_t$ limit is necessary to understand the impact of an enhanced $h\tilde t \tilde t^*$ interaction on the wave function $|R(0)|$. We will explore this interesting avenue in for future work.
\begin{table}[h!]
\begin{center}
\begin{tabular}{|c|c|c|c|c|}
\hline 
\multirow{2}{*}{Final state} & \multicolumn{2}{c|}{Signal Benchmark $S_{1}$} & \multicolumn{2}{c|}{Signal Benchmark $S_{2}$}\tabularnewline
\cline{2-5} \cline{3-5} \cline{4-5} \cline{5-5} 
 & \quad\, BR \quad \quad & $\Gamma$ {[}GeV{]} & \quad\, BR \quad \quad & $\Gamma$ {[}GeV{]}\tabularnewline
\hline 
\hline 
$g'g'$ & 0.63  & 0.078  & 0.33  & 0.078 \tabularnewline
\hline 
$HH$ & 0.31  & 0.033  & 0.61  & 0.133 \tabularnewline
\hline 
else & 0.06 & 0.007 & 0.06  & 0.014 \tabularnewline
\hline 
\end{tabular}
\end{center}
\caption{Estimate of the partial decay width of the stoponium state and comparison with the branching ratio for the two signal benchmarks highlighted in the right panel of Figs.~\ref{fig:moneyplot1}~and~\ref{fig:moneyplot2}. The ``else'' final state includes $WW,ZZ$ and $t\bar t$.}
\label{tab:BR}
\end{table}

\section{Final Remarks}
\label{s.conclusions}

We showed how di-Higgs resonances are predicted in Folded SUSY in the limit of large $A_t$, the parameter of the trilinear soft SUSY breaking term, in the stop sector. Our results are relevant for subsequent runs at the LHC, where these resonances could be discovered in the range of 500 - 1600 GeV under reasonable assumptions. These values correspond to stop masses between 250 and 800 GeV.

The observation that stoponium bound states preferentially decay to HH has been made in past in the context of the MSSM. However, these bound states can only be conceived in the MSSM for light stops, in a range of masses excluded by LHC searches. Our analysis brings back the possibility that stoponium bound states will produce HH resonances that the LHC will soon discover but this time in the context of F-SUSY. This makes a direct connection between HH resonances, the third generation of (s)quarks, and Naturalness.

Although our analysis focuses on F-SUSY, we argue that the main ingredients of the model that led us to the main results are also present in other models of NN. In general, in NN models the Higgs is the portal between the SM and the {\it dark} (or {\it mirror}) sectors. What we showed in this paper is that enhancing the parameter that connects the Higgs with the third generation quirks in the dark sector has two effects: it enhances the production of the corresponding squirkonium state, and it enhances its BR to HH. Once the LHC discovers a HH resonance, a thorough study of its decay modes will serve to unveil the underline theory responsible for said resonance. A pattern like the one in the right panel of Fig. \ref{fig:xsec} will be a smoking gun pointing at F-SUSY, and it will help us determine some of the model parameters. In a different model of NN the squirkonium bound state will have a different pattern of decays that deserve detailed study in future work.

\vspace{7mm}
\textbf{Acknowledgements:} 
We thank Zackaria Chacko for valuable discussions at the early stage of this work. We also thank David Curtin and Brian Batell for providing valuable feedback that helped improved this work. This manuscript has been authored by Fermi Research Alliance, LLC under Contract No. DE-AC02-07CH11359 with the U.S. Department of Energy, Office of High Energy Physics. MB is grateful to CNPq for the financial support.

\bibliographystyle{JHEP}
\bibliography{References}

\end{document}